\documentclass[3p,times,twocolumn]{elsarticle}

\usepackage{ecrc}
\usepackage{amsmath}
\usepackage{amssymb}
\volume{00}

\firstpage{1}

\journalname{Nuclear Physics B Proceedings Supplement}

\runauth{}

\jid{nppp}

\jnltitlelogo{Nuclear Physics B Proceedings Supplement}

\usepackage{amssymb}
\usepackage[figuresright]{rotating}

\def\sss{\scriptscriptstyle}

\def \bx {\mathbf{x}}

\newcommand{\Tint}[1]{{\hbox{$\sum$}\!\!\!\!\!\!\!\int\,}_{\!\!\!\!\raise-0.9ex\hbox{$\scriptstyle{#1}$}}}

\def\siml{{\ \lower-1.2pt\vbox{\hbox{\rlap{$<$}\lower6pt\vbox{\hbox{$\sim$}}}}\ }}
\def\simg{{\ \lower-1.2pt\vbox{\hbox{\rlap{$>$}\lower6pt\vbox{\hbox{$\sim$}}}}\ }}

\def \bx {\mathbf{x}}

\def \als {\alpha_{\mathrm{s}}}

\def \m2   {\mu^{2 \epsilon}}

\def\siml{{\ \lower-1.2pt\vbox{\hbox{\rlap{$<$}\lower6pt\vbox{\hbox{$\sim$}}}}\ }}
\def\simg{{\ \lower-1.2pt\vbox{\hbox{\rlap{$>$}\lower6pt\vbox{\hbox{$\sim$}}}}\ }}

\def\nn {\nonumber}

\def\mm {m_\infty^2}

\def\2to2{{2\leftrightarrow 2}}

\def\k{{\bf{k}}}

\def\x{{\bf{x}}}
\def\OO{{\mathcal{O}}}
\def\twotwo{{2\lra2}}
\def\lra{\leftrightarrow}
\def\nfd{n_{\!\sss F}}
\def\nbe{n_{\!\sss B}}

\begin{document}

\begin{frontmatter}



\dochead{}

\title{The thermal dilepton rate at NLO at small and large invariant mass}


\author{Jacopo Ghiglieri}

\address{Institute for Theoretical Physics, Albert Einstein Center, University of Bern,\\
Sidlerstrasse 5, CH-3012, Bern, Switzerland}

\begin{abstract}We report on a recent next-to-leading order perturbative determination of 
the dilepton rate from a hot QCD plasma for frequency and momentum of the order of
the temperature and for much smaller invariant mass $M\sim gT$. 
We briefly review the calculation, which generalizes the previous one for the photon case ($M=0$). 
We then analyze the consequences of the new calculation for the extraction of the photon 
rate from the small mass dilepton measurements. We then review a recent NLO determination
at large $M$ and we show how to match and merge its results with the low-mass ones, resulting
in a single rate which is NLO-accurate over the phenomenologically relevant region.
\end{abstract}

\begin{keyword}
	Dileptons, Hard Probes, Quark-Gluon Plasma, High order calculations, Lattice QCD


\end{keyword}

\end{frontmatter}

% \linenumbers

\section{Introduction}
\label{intro}
Electromagnetic (EM) probes have long been considered a key \emph{hard probe}
of the medium produced in ultrarelativistic heavy-ion collisions.  
Their chief advantage is that they are weakly coupled to the plasma, 
so that their reinteractions with it can be considered negligible. 
EM probes hence carry direct information about their formation process to the detectors,
unmodified by hadronization or other late time physics. 

In this contribution we will concentrate on dileptons. Compared to photons,
the kinematics of dileptons is described by two parameters, the frequency $k^0$
and the momentum $k$, with the related invariant mass $M\equiv\sqrt{k_0^2-k^2}$.
From an experimental point of view,
dileptons, compared to photons, have the advantage of a smaller background from meson decays, which
needs to be subtracted. For this reason, experimentalists have also focused on small-mass
dileptons, which can be thought of as massive off-shell photons.
Provided the mass of the pair is above the pion mass, the
pion decay background is absent and the foreground rates are
under much better control.  For this reason, $e^+ e^-$ pairs
with $M$ somewhat above $m_\pi^2$ have been measured, to
serve as an \textsl{ersatz} photon rate measurement
\cite{Adare:2008ab,Adare:2009qk,Adare:2015ila}.

In this contribution we will then first illustrate a recent 
perturbative calculation of the thermal dilepton rate at small $M$ (and for $k\sim T$)
to NLO \cite{Ghiglieri:2014kma},
extending the previous work on real photons \cite{Ghiglieri:2013gia}, aiming also
at understanding whether the rate, as a function of $M$, is smooth enough in going
from $M=0$ to finite $M$, so that the ersatz photon rate measurements are meaningful.
We will afterwards show the results of an NLO calculation at larger $M$ \cite{Laine:2013vma}
and then show how the small- and large-$M$ computations can be merged 
\cite{Ghisoiu:2014mha,Ghiglieri:2014kma}, resulting in a rate
that is reliably NLO for most invariant masses.  Another motivation for these NLO calculation
is to assess the reliability of the pQCD rates, widely employed in phenomenological analyses,
when extrapolated to $\als\sim 0.3$ where the coupling $g$ is not small. 
We will then conclude by remarking on the implications of the results on this matter, 
with an outlook to comparisons with non-perturbative lattice data. In all cases
the starting point is the formula giving the dilepton 
production rate per unit phase space 
at leading order in QED (in $\alpha$) and to all orders in QCD.
It reads (see for instance Ref.~\cite{LeBellac})
\begin{equation}
\label{defratedilepton}
\frac{d\Gamma_{l\bar l}}{d^4K}=-\frac{2\alpha}{3(2\pi)^4K^2}	
W^<(K) \, \theta( (k^0)^2 - \k^2) \,,
\end{equation}
where $K^2=(k^0)^2-k^2=M^2$ is the virtuality of the dilepton pair,
assumed much greater than $4m_l^2$. The rate is given in 
terms of the photon polarization $W^<(K)$, which reads
\begin{equation}
\label{defPi}
W^<(K) \equiv \int d^4X e^{i K\cdot X}
   \mathrm{Tr}\rho J^\mu(0)J_\mu(X)\,.
\end{equation}
Here $J^\mu=\sum_{q=uds}e_q\bar q\gamma^\mu q$ is the EM current
and we work in thermal equilibrium, so that the $\rho=e^{-\beta H}$
and the Hilbert space trace becomes a thermal average.\footnote{A
	calculation in an off-equilibrium setting relevant for heavy-ion
	collisions has been presented at this conference in \cite{Bhattacharya:2015wca}.} 
We will work perturbatively in the strong
coupling $g$, meaning that we treat the scale $gT$ 
(the \emph{soft scale}) as parametrically smaller than the scale $T$ 
(the \emph{hard scale}).
%other related theory talks if space at the end

\section{NLO at small $M$}
\label{sec_small}
\begin{figure}[ht]
	\begin{center}
		\includegraphics[width=6cm]{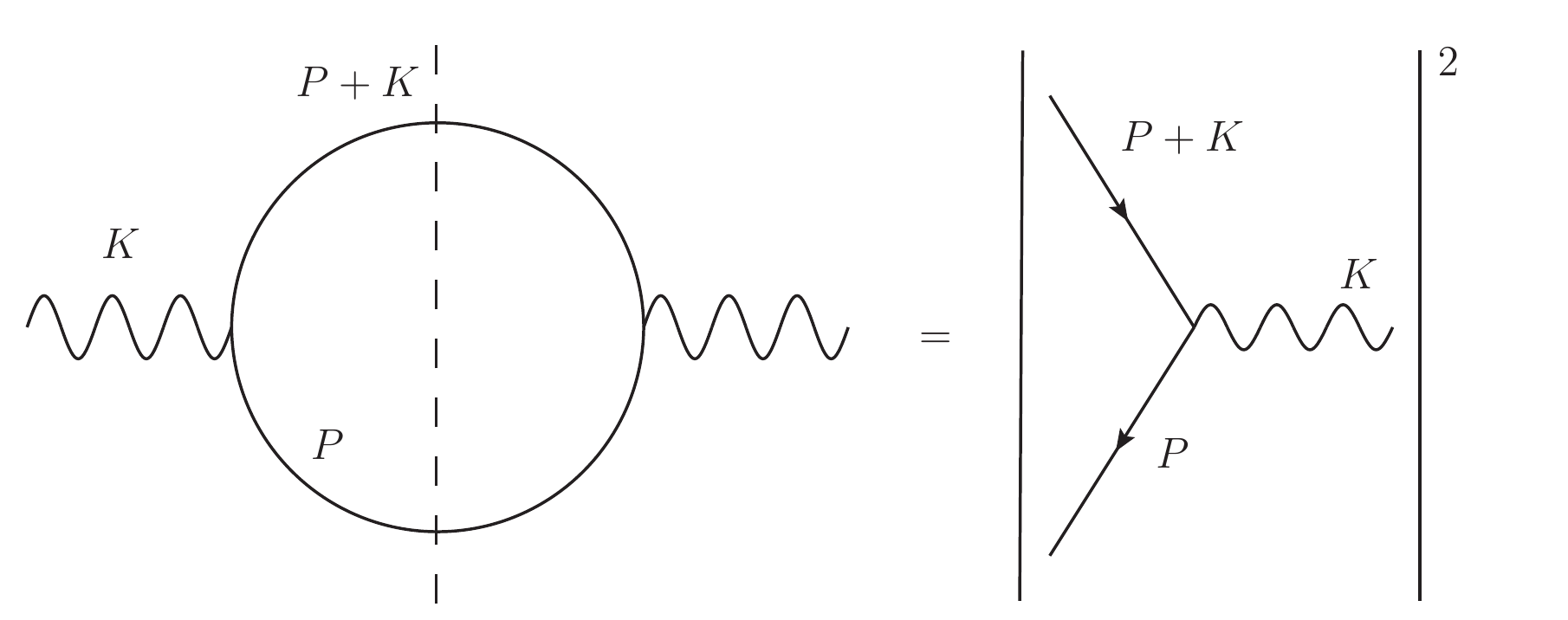}
	\end{center}
	\vspace{-0.3cm}
	\caption{The Born diagram on the left and the cut corresponding to the
		squared amplitude for the \emph{thermal Drell-Yan process} on the right.
	The plain lines with arrows are quarks and the photon is understood to be virtual;
its decay in the dilepton is not shown.}
	\label{fig_born}
\end{figure}

\begin{figure}[ht]
	\begin{center}
		\includegraphics[width=6cm]{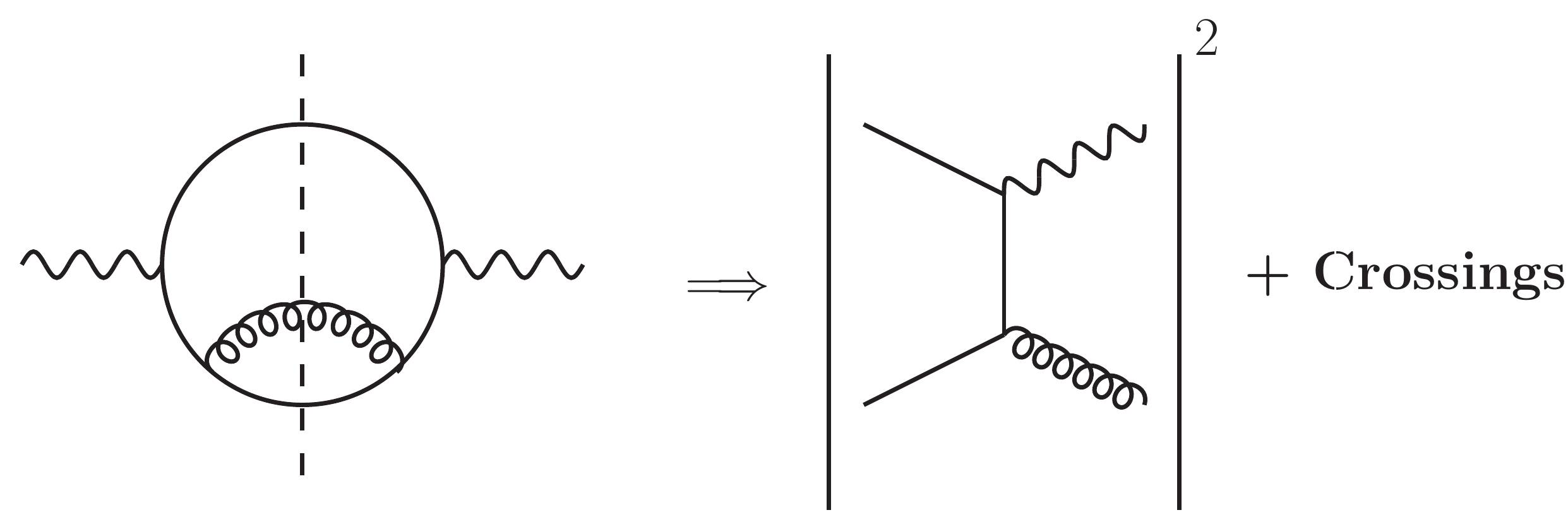}
	\end{center}
	\vspace{-0.3cm}
	\caption{$\twotwo$ processes. $1\lra3$ processes can be shown to be suppressed
	for $M\sim gT$.}
\label{fig_22}
\end{figure}

\begin{figure}[ht]
	\begin{center}
		\includegraphics[width=6cm]{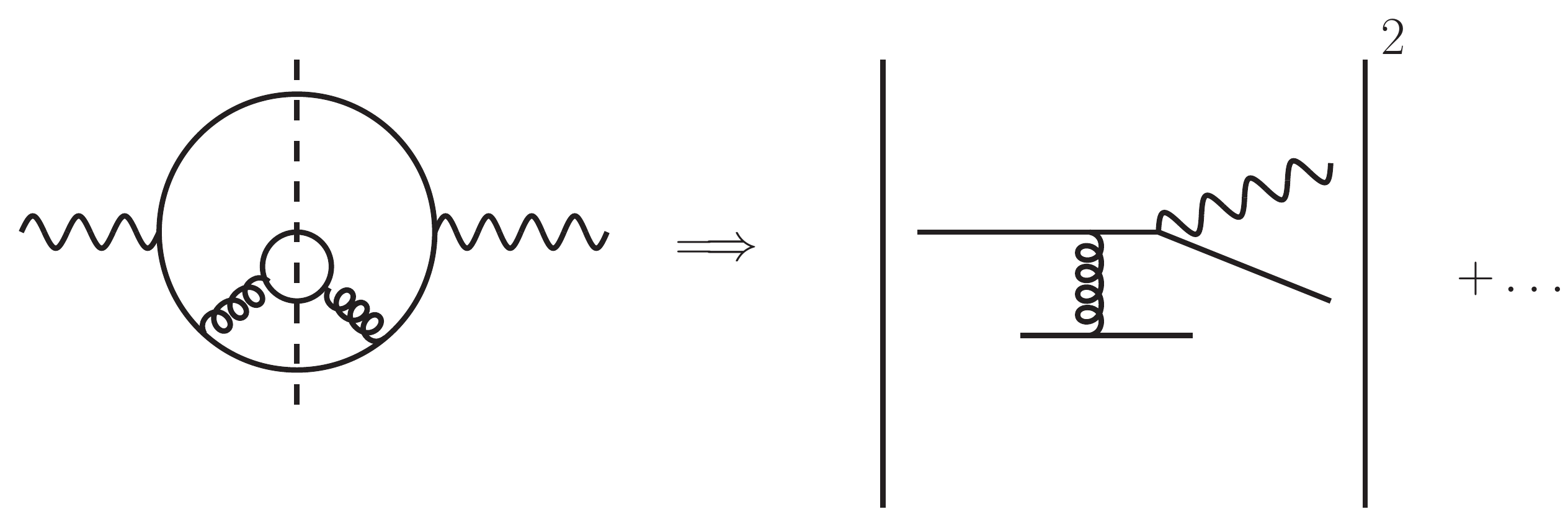}
	\end{center}
	\vspace{-0.3cm}
	\caption{Collinear processes.}
\label{fig_coll}
\end{figure}

By small $M$ we mean $M\sim gT$ (and $k\sim T$), so that, as we will show, the calculation
shares many similarities with the one for real photons \cite{Ghiglieri:2013gia}.
In a naive perturbative expansion the leading order term would be the \emph{Born term},
corresponding to the amplitude of the simple diagram shown in Fig.~\ref{fig_born}.
However, its contribution to $W^<$ scales approximately like $M^2$, so that
in our case its contribution is suppressed and other processes, apparently of
higher loop order, contribute at the same (leading) order in $g$. These are the
\emph{$\twotwo$ processes} shown in Fig.~\ref{fig_22} and the collinear processes
shown in Fig.~\ref{fig_coll}.
The former require some care when the $t$ or $u$ channel exchanged quark
become soft: the resulting logarthmic divergence is cured by Hard Thermal Loop
resummation \cite{Baier:1991em,Kapusta:1991qp}. 
At these small virtualities the calculation is unmodified w.r.t. the
real photon one. \\
Collinear processes are apparently suppressed w.r.t. the $\twotwo$ ones. However,
they receive an enhancement when the quark and antiquark (in the annihilation case)
or the outgoing quark and the photon (in the bremsstrahlung case) are collinear.
Furthermore, the soft scatterings that induce the splitting/annihilation are so frequent
that, within the photon's formation time, many of them can occur and interfere, in what is called
the Landau-Pomeranchuk-Migdal (LPM) effect. Its treatment require the resummation of an infinite
number of ladder exchanges of spacelike HTL gluons. This has been done for photons first \cite{Arnold:2001ba}
and later extended to small-$M$ dileptons \cite{Aurenche:2002wq}. In the latter case it is important
to note that the Born term in Fig.~\ref{fig_born} is the zeroth-order term in the ladder resummation 
series, and is thus included in the treatment of the LPM effect.

In summary, the leading-order result can be written as\footnote{for QCD with $uds$ light quarks}
\begin{align}
	\nn W^<(K)_{\mathrm{LO}}=\frac{8\alpha_\mathrm{EM}\nfd(k)g^2T^2}{3}
	\bigg[&\ln\left( \frac{T}{m_\infty}\right)+C_{\twotwo}\left(\frac kT\right)\\
	&+C_\mathrm{coll}\left(\frac kT,\frac{M}{gT}\right) \bigg],
	\label{lo}
\end{align}
where the logarithm comes from the screening of the aforementioned divergence and
$\mm=g^2T^2/3$ is the thermal mass of quarks. $C_{\twotwo}$ is the coupling-independent
part of the $\twotwo$ processes.

At NLO both processes receive $\OO(g)$ corrections: the soft end of the $\twotwo$ region
is sensitive to the addition of one extra soft gluon and similarly the collinear sector
requires the resummation of  soft one-loop corrections to the ladder resummation. Furthermore
a new process, the \emph{semi-collinear} one, contributes. It can be seen as the next order
in a collinear expansion, where the angle is allowed to be a bit larger. It thus interpolates
between the $\twotwo$ and collinear limits. The calculation of all of these corrections requires
dealing with loops of soft HTL excitations, which are known to be a computational challenge, resulting
in multidimensional numerical integrals over the intricate HTL propagators and vertices. However
it has recently been found \cite{CaronHuot:2008ni,Ghiglieri:2013gia,Ghiglieri:2015zma} that such
calculations simplify tremendously when the related operators are at light-like separations, as is the
case for photons and small-mass dileptons. Furthermore, the evaluation of
soft and semi-collinear corrections for low-mass dileptons is unaffected by $M$ and thus
identical to the corresponding photon case. Only the NLO LPM resummation needs to be modified in order
to obtain the small-$M$ NLO correction. %, which reads
%\begin{align}
%	\nn \delta W(K)=\frac{8\alpha_\mathrm{EM}\nfd(k)g^2T^2}{3}
%	\bigg[&-g\sqrt{\frac{6}{\pi^2}}\ln\left( \frac{\sqrt{2 m_DT}}{m_\infty}\right)-g\sqrt{\frac{6}{\pi^2}}C_\mathrm{soft+sc}
%		\left(\frac kT
%		\right)\\
%	&+\delta C_\mathrm{coll}\left(\frac kT,\frac{M}{gT}\right) \bigg],
%	\label{nlo}
%\end{align}
Numerical results will be shown in Sec.~\ref{sec_res}.

%Basically, the hard modes that will eventually give rise
%to the (real or virtual) photon are travelling so fast that the soft background can't keep up with them, so that
%they only probe statistical rather than dynamical correlations. 

\section{Merging small and large $M$}
\label{sec_merge}
\begin{figure}[ht]
	\begin{center}
		\includegraphics[width=8cm]{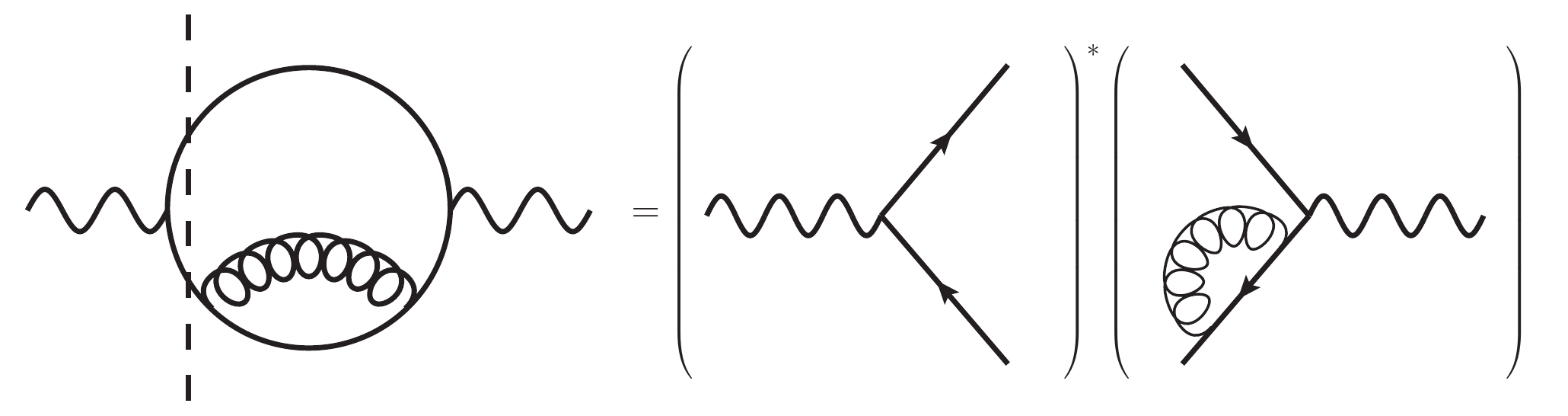}
	\end{center}
	\vspace{-0.3cm}
	\caption{A cut in a two-loop diagram for $W^<(K)$ corresponding to a
	virtual correction to the Born term.}
\label{fig_virt}
\end{figure}

An NLO perturbative calculation for $M\sim T$, $k\sim T$ has been presented in
\cite{Laine:2013vma}. In this region the Born term in Fig.~\ref{fig_born}
is the leading order and NLO is given by the $\twotwo$ processes in Fig.~\ref{fig_22},
together with their $1\lra3$ crossings, now kinematically allowed, and with
virtual corrections to the Born term, as shown in Fig.~\ref{fig_virt}. 
The NLO evaluation is
rather intricate, as one has to deal with the complicated kinematics of these processes
and with the fact that the virtual corrections to the Born term and the real corrections 
($\twotwo$ and $1\lra3$ processes) are separately IR divergent, requiring intermediate
regularizations. After the kinematics and divergences have been taken care of,
one is left with a set of two-dimensional numerical integrals.

This large-$M$ calculation does not require HTL resummation; hence, 
it diverges logarithmically for small $M$. Conversely, the small-$M$ calculation
described before behaves like $W^<\sim g^2(M^2+T^2)$ at large $M$, whereas an OPE 
analysis \cite{CaronHuot:2009ns} shows that no $T^2$-proportional term can exist for $M\gg T$. 
These drawbacks of the two calculations can be overcome by merging them in a single
one which has the right behaviour both at small and large $M$. This can be done
\cite{Ghisoiu:2014mha,Ghiglieri:2014kma} by taking the collinear part of the 
small $M$ calculation and expanding it for large $M$: the first and second term
in that expansion correspond to terms that are already included in the large $M$
calculation. Thus, if the collinear part of the small $M$ calculation
is added to the large-$M$ one, \emph{minus} these two terms, the result
is the sought after merged calculation. In \cite{Ghisoiu:2014mha}
it was presented with the LO collinear part and in \cite{Ghiglieri:2014kma} the small-$M$ NLO
corrections were added, leading to a merged result that is NLO-correct for $M\sim gT$, $M\sim \sqrt{g}T$
and $M\sim T$.

\section{Results and discussion}
\label{sec_res}
\begin{figure}[ht]
	\vspace{-0.4cm}
	\begin{center}
		\includegraphics[width=8cm]{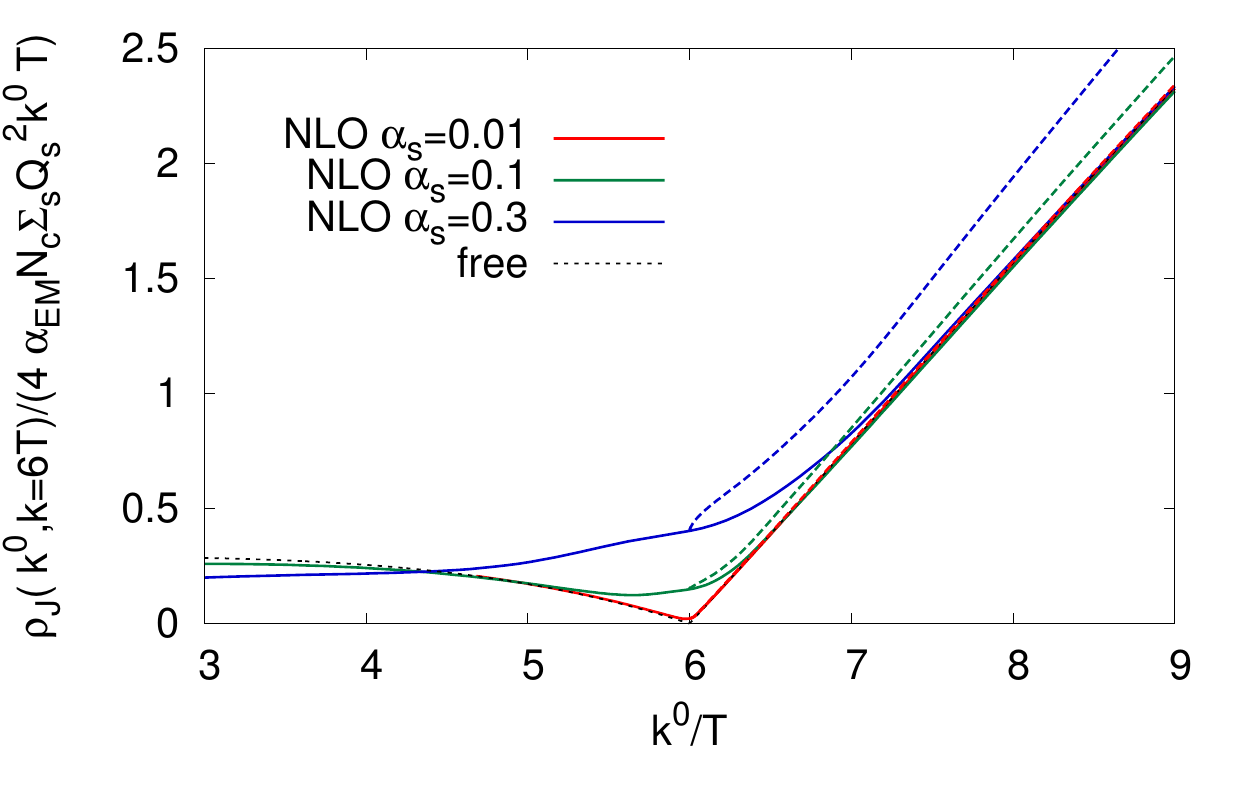}
	\end{center}
	\vspace{-0.8cm}
	\caption{The spectral function $\rho_J$ ($W^<=\nbe(k^0)\rho_J$) at NLO
	for three different values of the coupling and for $k=6T$. The solid lines are the 
	small-$M$ results \cite{Ghiglieri:2014kma}, which become unreliable away from the
light cone. The dashed line is the merge of the small- \cite{Ghiglieri:2014kma} and
large-$M$ results \cite{Laine:2013vma}, according to the procedure of \cite{Ghisoiu:2014mha,Ghiglieri:2014kma}.
Since no large-$M$ calculation is available for the spacelike region, no dashed line is shown there.
Picture taken from \cite{Ghiglieri:2014kma}. The dotted line is the free ($g=0$) spectral function. }
\label{fig_spf}
\end{figure}

The results of the aforementioned procedure are available online \cite{dilweb}\footnote{
	The code used to obtain the collinear part of the rates at LO and NLO is available
	in the arXiv submission for Ref.~\cite{Ghiglieri:2014kma}.}
and shown in Fig.~\ref{fig_spf}, where the spectral function associated to
$W^<(K)$ is plotted for different values of the coupling. The solid lines are the small-$M$ results only and the
dashed ones are the merged ones. At the smallest coupling the spectral function varies rapidly across the 
light cone, whereas the extrapolation at larger couplings shows a much smoother behaviour\footnote{The dashed
	are less smooth at the lightcone. However, as remarked in \cite{Ghisoiu:2014mha,Ghiglieri:2014kma},
the more reliable calculation there is the one in solid lines.}. This then gives more support to the
use of small-$M$ dileptons as a proxy for real photons.

\begin{figure}[ht]
	\vspace{-0.8cm}
	\begin{center}
		\includegraphics[width=8cm]{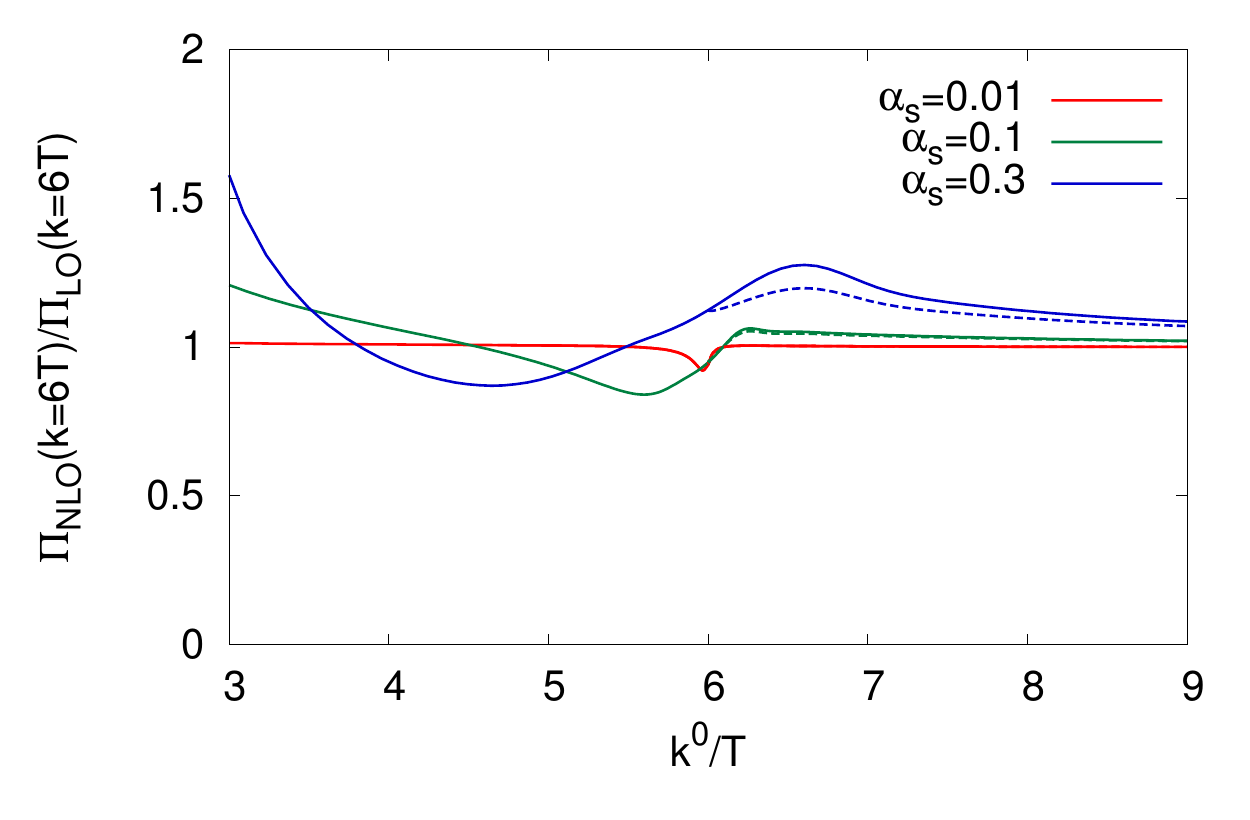}
	\end{center}
	\vspace{-0.8cm}
	\caption{The ratio of the NLO rates with respect to the LO ones
	for three different values of the coupling and for $k=6T$. The solid lines are the 
	small-$M$ results \cite{Ghiglieri:2014kma}, where the LO is given by Eq.~\eqref{lo}.
The dashed line is the ratio of the NLO merged results \cite{Ghiglieri:2014kma} shown in dashes
in Fig.~\ref{fig_spf} over the LO merged results \cite{Ghisoiu:2014mha}, which do not include
the small-$M$ NLO results described in Sec.~\ref{sec_small}.
Picture taken from \cite{Ghiglieri:2014kma}.}
\label{fig_ratio}
\end{figure}

One might still wonder, however, how reliable these perturbative calculations are, in particular
when extrapolated to $\als=0.3$ where $g$ is not small. Fig.~\ref{fig_ratio} shows that the NLO
corrections represent at most a 30\% increase, suggesting that the dilepton rate, similarly to the
photon one \cite{Ghiglieri:2013gia} is not plagued by the severe convergence problems affecting
other observables in thermal pQCD.

A further assessment of the reliability of these calculation could come from
an interplay of perturbative and non-perturbative inputs. As reported elsewhere
in this conference \cite{francis}, the direct extraction of the rates from
lattice calculations is extremely tricky: the lQCD can only access the
\emph{Euclidean} $JJ$ correlator, which is related to the spectral function as
\begin{align}
\nn W_E(\tau,\k)\equiv &
\int d^3x \left\langle J_\mu(\tau,\bx)J_\mu(0)\right\rangle e^{i\k\cdot\bx}\\
=&\int_0^\infty\frac{dk^0}{2\pi}\rho_J(k^0,\k)\frac{\cosh\left(k^0(\tau-1/(2T))\right)}{\sinh\left(\frac{k^0}{2T}\right)}.
\label{spf}
\end{align}
The inversion of this convolution is an ill-defined problem. On the other
hand, the continuation of the pQCD data to Euclidean spacetime is straightforward
\footnote{The Euclidean version of the results of \cite{Laine:2013vma,Ghisoiu:2014mha,
Ghiglieri:2014kma} is available online \cite{euclweb}, with parameters tuned to the 
lattice data in \cite{Ding:2014dua}. A zero-momentum dataset is also present, combining
data from \cite{Altherr:1989jc,Moore:2006qn}.}
and can be used for comparisons with lattice data, as attempted in 
\cite{Laine:2013vma}. As observed there, some care is necessary, as most
of the Euclidean correlator comes from the $k^0\gg k$ region of the spectral
function, which is dominated by well-understood vacuum physics, so that it might
not be easy to disentangle the contribution from the more interesting $k^0\sim k$
region.

\section{Conclusions}
We have shown how pQCD calculations are now available at NLO for the dilepton rate at
finite $k$ in a wide kinematical range. After briefly reviewing the intricacies
involved in the determination of the small and large $M$ rates, we have shown
how the two can be merged in a single set, which is plotted in Fig.~\ref{fig_spf}
and collected online for phenomenological use in \cite{dilweb}\footnote{A first
	phenomenological analysis, coupling those rates to the hydrodynamical evolution
	of the medium, has been reported in \cite{Burnier:2015rka}.}.
	The rates appear to be only mildly affected by NLO corrections even
	for $\als=0.3$ and for these couplings are smooth across the light cone,
	giving support to the use of low-mass dileptons as an ersatz real photon measurement.\\
\emph{Acknowledgements} I thank G.~Moore and M.~Laine for collaboration.
	My work is supported by the Swiss
National Science Foundation (SNF) under grant 200020\_155935.


%% References

%\nocite{*}
\bibliographystyle{elsarticle-num}
\bibliography{Ghiglieri_J}






% \end{thebibliography}

\end{document}